\documentclass[conference]{./IEEEtran_modified}
\usepackage{graphicx}
\usepackage{enumerate}
\usepackage{enumitem}

\usepackage{cite}

\ifCLASSINFOpdf
\else
\fi
\usepackage[table,xcdraw]{xcolor}
\usepackage{url}

\usepackage{tikz}
\usepackage{textcomp}
\usepackage{hyperref}
\usepackage{lipsum}

\newcommand\copyrighttext{%
	\footnotesize \textcopyright 2018 IEEE. Personal use of this material is permitted. Permission from IEEE must be obtained for all other uses, in any current or future media, including reprinting/republishing this material for advertising or promotional purposes, creating new collective works, for resale or redistribution to servers or lists, or reuse of any copyrighted component of this work in other works. Accepted to be Published in: Proceedings of the 2018 IEEE International Conference on Software Maintenance and Evolution (ICSME), September 23-29, 2018, Madrid, Spain. 
}
\newcommand\copyrightnotice{%
	\begin{tikzpicture}[remember picture,overlay]
	\node[anchor=south,yshift=10pt] at (current page.south) {\fbox{\parbox{\dimexpr\textwidth-\fboxsep-\fboxrule\relax}{\copyrighttext}}};
	\end{tikzpicture}%
}

\begin{document}
%
\title{COBOL to Java and Newspapers Still Get Delivered}

\author{\IEEEauthorblockN{Alessandro De Marco}
\IEEEauthorblockA{The New York Times Company\\
New York, NY, USA\\
alex.demarco@nytimes.com}
\and
\IEEEauthorblockN{Valentin Iancu}
\IEEEauthorblockA{Modern Systems International Ltd.\\
Bucharest, Romania\\
viancu@modernsystems.com}
\and
\IEEEauthorblockN{Ira Asinofsky}
\IEEEauthorblockA{The New York Times Company (Retired)\\
New York, NY, USA\\
irabevasi@aol.com}}


%


\maketitle
\copyrightnotice

\begin{abstract}
This paper is an experience report on migrating an American newspaper company's business-critical IBM mainframe application to Linux servers by automatically translating the application's source code from COBOL to Java and converting the mainframe data store from VSAM KSDS files to an Oracle relational database. The mainframe application had supported daily home delivery of the newspaper since 1979. It was in need of modernization in order to increase interoperability and enable future convergence with newer enterprise systems as well as to reduce operating costs. Testing the modernized application proved to be the most vexing area of work. This paper explains the process that was employed to test functional equivalence between the legacy and modernized applications, the main testing challenges, and lessons learned after having operated and maintained the modernized application in production over the last eight months. The goal of delivering a functionally equivalent system was achieved, but problems remained to be solved related to new feature development, business domain knowledge transfer, and recruiting new software engineers to work on the modernized application.
\end{abstract}


%
\IEEEpeerreviewmaketitle

\section{Introduction}

Since 1979, this American news media company relied on a software application running on its mainframe as the core IT system supporting daily home delivery of its newspaper. The application had grown to more than two million lines of COBOL code implementing billing, customer account maintenance, delivery routing, and other business-critical functionality. As a legacy system, it ``represent[ed] years of accumulated experience and knowledge'' \cite{bennett1995legacy}, while it ``significantly resist[ed] modification and evolution'' \cite{brodie1995legacy}. It was also very expensive to operate in comparison to more modern systems at the company.

An attempt to redevelop the home delivery application between 2006 and 2009 failed. In 2015, with mounting pressure to quickly lower costs, a different modernization approach was selected. Instead of redeveloping the application from scratch, the aim was to migrate the application off the mainframe and onto Linux servers by using a code and data translation approach. This approach promised to deliver an application that would be functionally equivalent, cheaper to operate, and easier to integrate with in comparison to the original. An evaluation of alternate approaches determined that a second attempt at redeveloping the application would have been much more expensive, and rehosting \cite{microfocus} would have continued to lock-up data in proprietary technology. 

A vendor provided the technology to convert the code and data \cite{bphx2014}. Based on an early \textit{proof-of-concept} trial of the vendor's technology, it became clear that even though the translated software could work as expected it would likely be more costly to maintain and enhance than \textit{handcrafted} Java software. It would also require knowledge of COBOL programming idioms and mainframe concepts that most Java software engineers would not possess. Despite the disadvantages, at an estimated cost of less than a tenth of the 2006-2009 redevelopment initiative and with a projected timeline of just one year, senior IT management opted to move forward.

As the project team leaders, we report on what went well, what did not, and lessons learned. Even though the modernized application went live a year later than originally planned, this is a success story.

There are known challenges inherent in legacy code translation \cite{terekhov2000realities}. However, in our project, testing the application proved to be the most time-consuming, difficult, and underestimated area of work. Unexpected testing obstacles caused significant delays. Development of an elaborate testing process was required in order to test functional equivalence between the legacy application and the modernized application, as the terms are defined in \cite{khadka2015does}, while supporting some functionality changes and feature enhancements along the way. 

\smallskip
The main contributions of this paper are:

\begin{enumerate}

\item Our process to test functional equivalence between the legacy and modernized applications.

\item The obstacles that we encountered while testing the modernized application.

\item Issues encountered in production due to gaps in the testing process and other lessons learned.

\end{enumerate}

In \ref{methodology}, we describe our migration methodology. In \ref{testing_process}, we explain our testing process. In \ref{discussion}, we cover the main challenges encountered in the testing process, production issues, and lessons learned. We conclude in \ref{conclusion}.

\section{Migration Methodology} \label{methodology}

\subsection{An 8-Step Process} \label{methodology_8steps}

We employed an 8-Step process to migrate our application off the mainframe and onto Linux servers. The steps are listed below. For a detailed explanation of the steps, refer to \cite{bphx2014}.

\begingroup
\small\selectfont

\begin{enumerate}[label=Step \arabic*:,leftmargin=5em]
	\item Collect Inventory in the Legacy System
	\item Break down into Work Packets
	\item Database Remodelling	
	\item Data Migration
	\item Code / JCL Conversion	
	\item Testing	
	\item User Acceptance Testing
	\item Cutover
\end{enumerate}

\endgroup
	
Work Packets are defined as one or more component groups of the application that could be translated and tested together. For each Work Packet, we executed Steps 3-6. Once all Work Packets had been converted and tested, we executed Steps 7 and 8.

\subsection{Technology Conversion Mapping Summary}

A high-level technology conversion mapping is provided in the table below. A detailed explanation of the vendor's translation technology (i.e. first 4 rows) is available in \cite{bphx2014}.

\begin{center}
\resizebox{9cm}{!} {
  \begin{tabular}{ | l | l | l | }
    \hline
    \parbox[t]{2.5cm}{\textbf{Technology}} & \parbox[t]{2cm}{\textbf{Legacy}} & \parbox[t]{4.5cm}{\textbf{Modernized}} \\ \hline
    \parbox[t]{2.5cm}{Programming Language} & \parbox[t]{2cm}{COBOL} & \parbox[t]{4.5cm}{Java}  \\ \hline
    \parbox[t]{2.5cm}{Database} & \parbox[t]{2.5cm}{VSAM KSDS files} & \parbox[t]{4.5cm}{Relational Database (Oracle)} \\ \hline
    \parbox[t]{2.5cm}{Batch Jobs} & \parbox[t]{2cm}{JCL} & \parbox[t]{4.5cm}{Spring Batch XML (JSR-352)} \\ \hline
    \parbox[t]{2.5cm}{User Interface Screens} & \parbox[t]{2cm}{BMS Maps} & \parbox[t]{4.5cm}{JSF, HTML/CSS, Javascript; accessible via a Web Browser} \\ \hline
    \parbox[t]{2.5cm}{Security and Access Control} & \parbox[t]{2cm}{RACF} & \parbox[t]{4.5cm}{Spring Security and Active Directory} \\ \hline
    \parbox[t]{2.5cm}{Middleware} & \parbox[t]{2cm}{CICS} & \parbox[t]{4.5cm}{Jetty, Apache-CXF, JPA/Eclipselink} \\ \hline
    \parbox[t]{2.5cm}{Reporting} & \parbox[t]{2cm}{QMF and DB2} & \parbox[t]{4.5cm}{Jasper Reports and Oracle} \\ \hline
    \parbox[t]{2.5cm}{Encryption} & \parbox[t]{2cm}{Megacryption} & \parbox[t]{4.5cm}{Java Cryptography Extension} \\ \hline
    \parbox[t]{2.5cm}{Screen Automation / Macros} & \parbox[t]{2cm}{IBM HATS} & \parbox[t]{4.5cm}{Newly developed Web Services} \\ \hline
    \parbox[t]{2.5cm}{Batch Process Scheduler} & \parbox[t]{2cm}{CA7} & \parbox[t]{4.5cm}{Control-M}  \\ \hline
    \parbox[t]{2.5cm}{Monitoring and Alerting} & \parbox[t]{2cm}{RMF, SMF, Omegamon} & \parbox[t]{4.5cm}{NewRelic, nagios, Sumologic}\\ \hline
    \parbox[t]{2.5cm}{Development and Deployment} & \parbox[t]{2cm}{Changeman} & \parbox[t]{4.5cm}{git, gradle, jenkins, puppet, ansible} \\ \hline

  \end{tabular}
}
\end{center}

\subsection{Framework Support For Utilities and Middleware Services}

In addition to translating the code and data, the vendor provided a runtime framework that implemented many mainframe services and utilities. This allowed the translated code to run on the modernized platform while continuing to interface with its environment in a similar way to how it did on the mainframe. This approach was also used in \cite{sneed2013migrating}.

\subsection{New Component Development}

When legacy application dependencies were not supported by the runtime framework (e.g. REXX, GVEXPORT), if an off-the-shelf software package was not available as a substitute, or if it made more sense to make use of capabilities of the new environment (e.g. Oracle database backups and restore points; file system snapshots), then replacement components were developed. In addition, since automatically converting the CA7 batch schedule for Control-M was unsuccessful, we redeveloped the batch job schedule for Control-M. 

\section{Testing Process} \label{testing_process}

To assure that the modernized application would be \textit{functionally equivalent} to the legacy application, we developed the testing process described next. By functional equivalence we mean that the modernized application would produce the same output as the legacy application given the same input. As the translation from COBOL to Java preserved business rules and much of the internal component hierarchy of the legacy application, we were able to test functional equivalence at the component group level first, and gradually build up to testing functional equivalence of the whole application.

\subsection{Black-Box Testing of Component Groups}

Component groups assembled related components that would be runnable and testable together as a ``black-box'' through existing externally accessible interfaces, such as SOAP Web Services, User Interface screens, database tables, and files. Given the same inputs, the output of legacy component groups were compared to the output of their modernized counterparts. When they matched, the test was considered to have passed. When they did not, root-cause analysis was performed to find the source of the mismatch.

\subsection{Test Environments}

A new test region \label{test_environments_mf_test_region} was created on the mainframe to support the testing process. It served as the \textit{source of truth} for expected behavior.

QA analysts and developers installed a Linux virtual machine and a database on their individual computers. This allowed everyone to test modernized components locally. 

Development, Staging, and Production environments were provisioned in a private datacenter. A shared Oracle database and a Control-M batch scheduler were configured in each. These environments let us build and test infrastructure and configuration automation code, the full batch schedule, and infrastructure-related performance improvements, since these could not be tested on individual developer machines. Static test data was used in the Development environment, and dynamic \textit{current day} test data was used in the Staging environment. To generate current day test data, the batch process ran every day in the mainframe test region and output files were then transferred over. This enabled testing of scenarios that depended on the day of week or month, and it validated that one day's batch output would be processed correctly the next day.

\subsection{Stages In the Testing Process}

The testing process, Step 6 of \ref{methodology_8steps}, was broken down into stages, with each stage progressively increasing in scope and level of difficulty in isolating the root-cause of test failures. The stages are summarized in the table below.

\begin{center}

	\resizebox{9cm}{!} {

		\begin{tabular}{ | l | l | }
			\hline
				\multicolumn{2}{|l|}{\textbf{Stage 1:} \textbf{Pre-Delivery}} \\ \hline
				\parbox[t]{0.5cm}{\textit{D:}} & \parbox[t]{8.5cm}{Tests done by the translation technology vendor prior to delivering translated code.} \\ 	\hline
				\parbox[t]{0.5cm}{\textit{E:}} & \parbox[t]{8.5cm}{Individual developer machines.} \\ \hline
				\parbox[t]{0.5cm}{\textit{S:}} & \parbox[t]{8.5cm}{One batch job, one set of screens, or a Web Service; automated testing of 10 batch jobs for regression testing.} \\ 
				\hline			
		\end{tabular}
	}
\end{center}

\begin{center}
	\resizebox{9cm}{!} {
		\begin{tabular}{ | l | l | }
			\hline
			\multicolumn{2}{|l|}{\textbf{Stage 2:} \textbf{Data Migration Validation}} \\ \hline
			\parbox[t]{0.5cm}{\textit{D:}} & \parbox[t]{8.5cm}{Tests done by the QA analysts to make sure that data loaded into the database matched VSAM KSDS files.}
			 \\ 	\hline
			\parbox[t]{0.5cm}{\textit{E:}} & \parbox[t]{8.5cm}{Individual QA analyst machines, and Development.} \\ \hline
			\parbox[t]{0.5cm}{\textit{S:}} & \parbox[t]{8.5cm}{All migrated data.} \\ 
			\hline			
		\end{tabular}
	}
\end{center}

\begin{center}
	\resizebox{9cm}{!} {
		\begin{tabular}{ | l | l | }
			\hline
			\multicolumn{2}{|l|}{\textbf{Stage 3:} \textbf{Component Group}} \\ \hline
			\parbox[t]{0.5cm}{\textit{D:}} & \parbox[t]{8.5cm}{Tests done by QA analysts to make sure that a component group worked as expected. }\\ 	\hline
			\parbox[t]{0.5cm}{\textit{E:}} & \parbox[t]{8.5cm}{Individual QA analyst machines, and, for Web Services and UI screens, both Development and Staging.} \\ \hline
			\parbox[t]{0.5cm}{\textit{S:}} & \parbox[t]{8.5cm}{One batch job, one or more related UI screens, one SOAP Web Service.} \\ 
			\hline			
		\end{tabular}
	}
\end{center}

\begin{center}
	\resizebox{9cm}{!} {
		\begin{tabular}{ | l | l | }
			\hline
			\multicolumn{2}{|l|}{\textbf{Stage 4:} \textbf{Batch Process (Operating on Static Test Data)}} \\ \hline
			\parbox[t]{0.5cm}{\textit{D:}} & \parbox[t]{8.5cm}{Automated tests for the end to end batch process. Also acted as a batch regression test system.} \\ 	\hline
			\parbox[t]{0.5cm}{\textit{E:}} & \parbox[t]{8.5cm}{Development} \\ \hline
			\parbox[t]{0.5cm}{\textit{S:}} & \parbox[t]{8.5cm}{The full batch, but configured to run the same day, everyday in order to simplify root-cause analysis.} \\ 
			\hline			
		\end{tabular}
	}
\end{center}
		
\begin{center}
	\resizebox{9cm}{!} {
		\begin{tabular}{ | l | l | }
			\hline
			\multicolumn{2}{|l|}{\textbf{Stage 5:} \textbf{Batch Process (Operating on Dynamic Test Data)}} \\ \hline
			\parbox[t]{0.5cm}{\textit{D:}} & \parbox[t]{8.5cm}{Automated tests for the end to end batch process.} \\ 	\hline
			\parbox[t]{0.5cm}{\textit{E:}} & \parbox[t]{8.5cm}{Mainframe test region and Staging, running in parallel.} \\ \hline
			\parbox[t]{0.5cm}{\textit{S:}} & \parbox[t]{8.5cm}{The full batch, but configured to compare current day output between the legacy and modernized processes.} \\ 
			\hline			
		\end{tabular}
	}
\end{center}

\begin{center}
	\resizebox{9cm}{!} {
		\begin{tabular}{ | l | l | }
			\hline
			\multicolumn{2}{|l|}{\textbf{Stage 6:} \textbf{Batch Process (Operating on Full Production Data)}} \\ \hline
			\parbox[t]{0.5cm}{\textit{D:}} & \parbox[t]{8.5cm}{Automated execution of the batch process in production to benchmark performance.} \\ 	\hline
			\parbox[t]{0.5cm}{\textit{E:}} & \parbox[t]{8.5cm}{Production (prior to cutover).} \\ \hline
			\parbox[t]{0.5cm}{\textit{S:}} & \parbox[t]{8.5cm}{The full batch, operating on production data.} \\ 
			\hline			
		\end{tabular}
	}
\end{center}

\begin{center}
	\resizebox{9cm}{!} {
		\begin{tabular}{ | l | l | }
			\hline
			\multicolumn{2}{|l|}{\textbf{Stage 7:} \textbf{System Integration}} \\ \hline
			\parbox[t]{0.5cm}{\textit{D:}} & \parbox[t]{8.5cm}{Tests done by QA analysts in collaboration with other teams/systems within the organization. } \\ 	\hline
			\parbox[t]{0.5cm}{\textit{E:}} & \parbox[t]{8.5cm}{Staging} \\ \hline
			\parbox[t]{0.5cm}{\textit{S:}} & \parbox[t]{8.5cm}{The whole enterprise system, with new transactions entered via client systems, processed by the batch, and flowing to downstream consumers via reports and file feeds.} \\ 
			\hline			
		\end{tabular}
	}
\end{center}

\centerline{\textit{D = Description, E = Environment, S = Scope}}

\section{Discussion} \label{discussion}

We discuss the main obstacles encountered in the Testing Process defined in \ref{testing_process}, types of production issues that were not caught in testing, and lessons learned.

\subsection{Testing Obstacles}

Testing accounted for approximately 70\% to 80\% of the time spent on the project. The project was completed a year later than expected due primarily to these obstacles. We have grouped the obstacles into three areas.

\subsubsection{Initial State of the Legacy Application} \label{initial_state}

\paragraph{Lack of Tests} The vast majority of legacy application components did not have a high enough level of test coverage to codify the required information for functional equivalence testing. As a consequence, much time was spent trying to analyze inputs, outputs, and sometimes even the internal behavior of components when tests failed and root-cause analysis had to be performed. This was especially time-consuming and difficult for the more complex batch jobs (i.e. many man-months of work). 

\paragraph{Lack of Batch Automation} The CA7 batch process scheduler had never been installed in the existing mainframe test region. In order to test functional equivalence of the batch process end-to-end, Control-M was installed and configured in a new test region as noted in \ref{test_environments_mf_test_region}. This work was unplanned and required a great amount of effort.

\paragraph{Obsolete Code} In operation for more than 35 years, the legacy application had accumulated a fair amount of obsolete code \cite{mdsy2017}. Due to a lack of adequate maintenance over the years, we spent time identifying this code and removing it to reduce the amount of translation and testing work to do.

\paragraph{Interfaces with Other Systems} The legacy application generated more than 3500 data file feeds and reports for downstream consumers daily. We did not know all the consumers, many transfers used insecure protocols, and connections were configured in a variety of different ways. Discovering the consumers, upgrading to secure protocols, and harmonizing configuration management caused significant delays.

\subsubsection{Data Formats} \label{data_formats}

\paragraph{EBCDIC Files That Contained Computational Fields}
Many files on the mainframe contained a mix of display fields and computation fields. The mainframe environment has appropriate tools for working with these file formats, but the Linux platform does not. Files of this kind had to be transferred over because they were required for functional equivalence testing of batch jobs. When transferring these files off the mainframe, the display fields needed to be converted from EBCDIC to ASCII, but the computational fields (e.g. COMP3) had to remain binary-encoded. ETL software was developed for this which required additional development work, became a processing bottleneck, and was a source of errors. Handling the large volume of files and many versions of each further complicated the matter.

\paragraph{File Processing Tools On Linux} A lack of tools to inspect, modify, and compare files with fixed block or variable block structures and computational fields on the Linux platform necessitated the development of new tools. This work was not part of the original plan. 

\subsubsection{Batch Performance} \label{batch_performance}

\paragraph{Multi-layout VSAM KSDS Files}

VSAM KSDS files with multiple record layouts (i.e. REDEFINEs) could be translated to one or many tables. We obtained better performance when mapping to one table, but improved maintainability (i.e. fewer null-value columns) when mapping each layout to its own table. Since reading records in an indexed VSAM file often involved moving a pointer to an indexed location given a key, and then iteratively moving the pointer forward or backward to the next or previous record until some condition was met, when translated to one relational database table using a result set cursor as the pointer, this usually performed well. When translated to 97 tables, the worst case we encountered, this was painfully slow because open cursors had to be maintained on result sets from each of the 97 tables, and then data access logic decided which cursor would have the next or previous record.

\paragraph{In-Memory VSAM Cache} \label{vsam_cache}

When testing some batch jobs, we encountered performance problems that could not be solved by database remodelling as described previously. The technology vendor developed an in-memory representation of VSAM KSDS files that acted like a cache. Encapsulated within data access middleware and tuned with external configuration settings, no application code needed to be changed. Once all necessary data had been loaded into memory, operations were performed against the in-memory data structure, and at the end of processing, changes were written back to the database. By performing these operations in-memory, network latency and database I/O bottlenecks were eliminated, but we could not run other batch jobs at the same time if they depended on the same database tables.

%
%

\subsection{Types of Issues That The Testing Process Failed to Uncover} \label{prd_issues}

Despite the heavy investment in testing, there were gaps in the testing process. After completing User Acceptance Testing and Cutover steps of \ref{methodology_8steps}, we encountered problems in production due to the gaps. We have grouped them by type with examples. 

\paragraph{Unexpected User Input} On the first day of operation, a subscriber contacted our call center to complain about non-delivery of the newspaper over the prior 147 days. Complaints over such a long time period are extremely uncommon. The algorithm to compute the credit due the subscriber involved a date computation that overflowed a field that supported a maximum of 99 days. This overflow caused a runtime exception and a critical batch job failed. When attempting to reproduce this on the mainframe, we observed that the date value was simply truncated causing the calculation to be incorrect, but the job did not fail.

\paragraph{Concurrency Issues} On another day, Control-M unfortunately scheduled two batch jobs to run at the same time. They both wrote to the same output file which corrupted the data in that file. The corrupted data was later loaded into an in-memory VSAM cache (see \ref{vsam_cache}) which caused a critical batch job to fail.

\paragraph{Inefficient Processing Idioms} Several batch jobs involved maintenance of data. On the mainframe, data maintenance was often done by transferring VSAM data to files which would then be pruned and sorted using file processing tools. The cleaned up files would then be reloaded into VSAM and reindexed. The translation of these batch jobs resulted in operations (i.e. millions of DELETEs and INSERTs) that locked up our database for hours on two different occasions causing online transaction processing outages. These maintenance jobs were redeveloped to make use of more efficient in-database maintenance operations or decommissioned altogether.

\subsection{Lessons Learned} \label{lessons}

\paragraph{Application Understanding}
As noted in a 2013 Gartner Survey cited in \cite{mdsy2017}, most modernization projects are delivered later than originally planned ``as a direct result of poor legacy application understanding''. The translation approach did not eliminate the need for us to develop a deep level of application understanding in critical parts of the system. We could have developed this knowledge while improving the initial state of the application as a precursor project to this one. This would likely have alleviated the problems noted in \ref{initial_state} and informed planning, estimation, and management of the project to follow.

\paragraph{Project Management} We attempted to break work down into smaller tasks and measure time taken to complete the tasks in order to forecast when similar future work would be completed, but this did not work well. Later Work Packets were more batch-oriented than earlier ones, so testing obstacles related to batch processing (\ref{data_formats} and \ref{batch_performance}) caused early delivery forecasts to be way off. This problem was discovered too late to remedy, and may have been avoided had we had a deeper understanding of the application when structuring the Work Packets at the outset.

\section{Conclusion} \label{conclusion}
We employed a code and data translation approach and an 8-step methodology to migrate a legacy application off the mainframe to Linux servers. We achieved the goal of delivering a modernized application that is functionally equivalent to the original. Functional equivalence was demonstrated via an elaborate testing process. It was delivered later than expected due to unexpected testing obstacles, but still at much lower cost than a prior redevelopment project that failed. Despite a few production issues, the modernized application has proven to be quite stable in production over the last eight months and newspapers still get delivered daily. 

On the other hand, new feature development remains challenging. It requires knowledge of COBOL programming idioms and mainframe concepts that Java software engineers on the team do not possess. Mainframe COBOL developers on the team know the idioms and concepts, but are not yet proficient in Java. We are cross-training developers and hope that this will help with knowledge transfer. We also note that it has been difficult to recruit new Java developers due to a lack of interest in working with the translated code.


%
%



\bibliography{ICSME}{}

\begin{thebibliography}{1}
\providecommand{\url}[1]{#1}
\csname url@samestyle\endcsname
\providecommand{\newblock}{\relax}
\providecommand{\bibinfo}[2]{#2}
\providecommand{\BIBentrySTDinterwordspacing}{\spaceskip=0pt\relax}
\providecommand{\BIBentryALTinterwordstretchfactor}{4}
\providecommand{\BIBentryALTinterwordspacing}{\spaceskip=\fontdimen2\font plus
\BIBentryALTinterwordstretchfactor\fontdimen3\font minus
  \fontdimen4\font\relax}
\providecommand{\BIBforeignlanguage}[2]{{%
\expandafter\ifx\csname l@#1\endcsname\relax
\typeout{** WARNING: IEEEtran.bst: No hyphenation pattern has been}%
\typeout{** loaded for the language `#1'. Using the pattern for}%
\typeout{** the default language instead.}%
\else
\language=\csname l@#1\endcsname
\fi
#2}}
\providecommand{\BIBdecl}{\relax}
\BIBdecl

\bibitem{bennett1995legacy}
K.~Bennett, ``Legacy systems: Coping with success,'' \emph{IEEE software},
  vol.~12, no.~1, pp. 19--23, 1995.

\bibitem{brodie1995legacy}
M.~L. Brodie and M.~Stonebraker, \emph{Legacy Information Systems Migration:
  Gateways, Interfaces, and the Incremental Approach}.\hskip 1em plus 0.5em
  minus 0.4em\relax Morgan Kaufmann Publishers Inc., 1995.

\bibitem{microfocus}
\BIBentryALTinterwordspacing
\relax{Micro Focus}, ``\relax{Micro Focus Enterprise Server Overview}.''
  [Online]. Available:
  \url{https://www.microfocus.com/products/enterprise-suite/enterprise-server/}
\BIBentrySTDinterwordspacing

\bibitem{bphx2014}
\BIBentryALTinterwordspacing
\relax{Modern Systems International Ltd., formerly BluePhoenix Solutions},
  ``\relax{Mainframe Migration COBOL to Java},'' 2014. [Online]. Available:
  \url{https://www.platformmodernization.org/bluephoenix/Lists/ResearchPapers/Attachments/2/BPHXApplication_Modernization_Whitepaper.pdf}
\BIBentrySTDinterwordspacing

\bibitem{terekhov2000realities}
A.~A. Terekhov and C.~Verhoef, ``The realities of language conversions,''
  \emph{IEEE Software}, vol.~17, no.~6, pp. 111--124, 2000.

\bibitem{khadka2015does}
R.~Khadka, P.~Shrestha, B.~Klein, A.~Saeidi, J.~Hage, S.~Jansen, E.~van Dis,
  and M.~Bruntink, ``Does software modernization deliver what it aimed for? a
  post modernization analysis of five software modernization case studies,'' in
  \emph{Software Maintenance and Evolution (ICSME), 2015 IEEE International
  Conference on}.\hskip 1em plus 0.5em minus 0.4em\relax IEEE, 2015, pp.
  477--486.

\bibitem{sneed2013migrating}
H.~M. Sneed and K.~Erdoes, ``\relax{Migrating AS400-COBOL to Java: a report
  from the field},'' in \emph{Software Maintenance and Reengineering (CSMR),
  2013 17th European Conference on}.\hskip 1em plus 0.5em minus 0.4em\relax
  IEEE, 2013, pp. 231--240.

\bibitem{mdsy2017}
\BIBentryALTinterwordspacing
\relax{Modern Systems International Ltd.}, ``\relax{Three Essentials for
  Reducing Risk When Migrating from COBOL to Java},'' 2017. [Online].
  Available: \url{http://modernsystems.com/essentials-cobol-to-java/}
\BIBentrySTDinterwordspacing

\end{thebibliography}
\bibliographystyle{IEEEtran}

%
%
%

\end{document}